\documentclass[runningheads]{llncs}
\usepackage[T1]{fontenc}
\usepackage{wasysym}
\usepackage{graphicx}
\usepackage{multirow}
\usepackage{xcolor}
\usepackage{amssymb}
\usepackage{amsfonts}
\usepackage{algorithm}
\usepackage{algpseudocode}

\usepackage{amsmath}
\usepackage{bm}
\usepackage{hyperref}
\usepackage{pdfpages}  
\usepackage{subcaption}

\newcommand{\sign}{\operatorname{sign}}
\newcommand{\cosSim}{\operatorname{cos}}

\begin{document}
\title{A Cognitive Distribution and Behavior-Consistent Framework for Black-Box Attacks on Recommender Systems}

\titlerunning{Black-Box Attacks on Recommender Systems}

\author{Hongyue Zhang\inst{1,2} \and Mingming Li\inst{1,2} \and Dongqin Liu \inst{1,2}$^*$ \and Hui Wang\inst{1,2} \and Yaning Zhang\inst{3} \and Xi Zhou\inst{1,2} \and Honglei Lv\inst{1,2}$^*$ \and Jiao Dai\inst{1,2} \and Jizhong Han\inst{1,2}}
\authorrunning{H.Zhang et al.}
%
\institute{Institute of Information Engineering, Chinese Academy of Sciences, China \and
School of Cyber Security, University of Chinese Academy of Sciences, China \and NSFOCUS, China\\
\email{\{zhanghongyue,limingming,liudongqin,wanghui4042,zhouxi,\\lvhonglei,daijiao,hanjizhong\}@iie.ac.cn}\\
\email{\{zhangyaning2\}@nsfocus.com}
}

\maketitle              

\let\thefootnote\relax\footnotetext{
$^*$ is the corresponding author.}
\begin{abstract}
With the growing deployment of sequential recommender systems in e-commerce and other fields, their black-box interfaces raise security concerns: models are vulnerable to extraction and subsequent adversarial manipulation. Existing black-box extraction attacks primarily rely on hard labels or pairwise learning, often ignoring the importance of ranking positions, which results in incomplete knowledge transfer. Moreover, adversarial sequences generated via pure gradient methods lack semantic consistency with real user behavior, making them easily detectable. To overcome these limitations, this paper proposes a dual-enhanced attack framework. First, drawing on primacy effects and position bias, we introduce a cognitive distribution-driven extraction mechanism that maps discrete rankings into continuous value distributions with position-aware decay, thereby advancing from order alignment to cognitive distribution alignment. Second, we design a behavior-aware noisy item generation strategy that jointly optimizes collaborative signals and gradient signals. This ensures both semantic coherence and statistical stealth while effectively promoting target item rankings. Extensive experiments on multiple datasets demonstrate that our approach significantly outperforms existing methods in both attack success rate and evasion rate, validating the value of integrating cognitive modeling and behavioral consistency for secure recommender systems.

\end{abstract}
\section{Introduction}
Currently, recommender systems have become a core component of information service platforms, widely deployed in diverse applications such as e-commerce \cite{liu2024unified,liMM,liqueryindex,li2023adaptive,lirerank}, short videos \cite{zhu2024interest}, and social media \cite{sharma2024survey}. Driven by the explosive growth of user behavior data, sequential recommendation models have significantly advanced personalized recommendation accuracy by effectively modeling the dynamic evolution of user interests \cite{boka2024survey}. Specifically, neural-structure-based models, including RNN and Transformers(e.g., BERT4Rec \cite{sun2019bert4rec}, SASRec \cite{kang2018self},
and NARM \cite{li2017neural}), excel at capturing complex temporal dependencies and contextual semantic relationships, making them widely adopted in industrial-grade recommender systems. However, the high complexity of the model and the exposed interfaces also bring new vulnerabilities, making them face complex black box extraction and adversarial manipulation threats \cite{yue2021black,zhu2023membership,zhao2025diversity}.

Existing research shows that even without access to the target model’s internal parameters or training data, attackers can replicate its behavior through a limited number of Application Programming Interface (API) queries \cite{zhao2025survey,zhao2025systematic}. Once a high-fidelity surrogate model is successfully constructed, it can be leveraged to craft adversarial perturbations that effectively mislead the original black-box system's recommendations \cite{lam2004shilling,pitsilis2019securing,zhang2021data,du2024precision}. While these findings deepen our understanding of recommendation system vulnerabilities, current black-box model extraction and attack mechanisms still suffer from two key limitations: (i)\textbf{ Model Extraction Deficiency}: Most existing black‑box recommendation attacks rely on hard labels or pairwise ranking learning frameworks \cite{yue2021black,wang2025sim4rec,zhu2023model,zhao2025llm4mea}. These approaches primarily utilize binary preference relations between items, overlooking the continuous value information conveyed by the relative importance of different ranking positions. This oversight leads to substantial information loss and restricts the surrogate model's ability to capture the fine-grained decision preferences of the black-box system. Furthermore, this simplification fails to account for the attention decay characteristics inherent in human cognition, thereby limiting both the theoretical upper bound and practical effectiveness of such attacks. (ii)\textbf{ Attack Stealth Compromise}: Pollution items generated by purely gradient‑based optimization methods often lack natural semantic relevance to the target item. This deficiency results in constructed user behavior sequences that exhibit statistical anomalies, rendering them easily detectable by existing defense mechanisms \cite{yue2021black}.

 To address these limitations, this paper proposes a dual-enhanced black-box attack framework. Our main innovations include:
(i)\textbf{ Cognitive Distribution-Driven Model Extraction}: Inspired by cognitive psychology's primacy effect \cite{asch1946forming,hogarth1992order} and position bias \cite{atsusaka2025analyzing}, we design a distribution reconstruction mechanism grounded in attention decay theory. This mechanism converts discrete ranking signals into continuous value distributions, thereby shifting the extraction paradigm from mere order alignment to cognitive distribution alignment.
(ii) \textbf{Behavior-Consistent Pollution Item Generation}: For attack generation, we introduce a pollution item selection strategy based on behavioral pattern consistency. By integrating collaborative relation analysis with gradient optimization, we develop a hybrid mechanism that jointly optimizes two complementary signals. This approach first mines historically co‑occurring items as candidate pollution entries, thus preserving semantic coherence and logical consistency in user behavior sequences. Subsequently, it leverages gradient information to maximize the impact on target item ranking. The resulting strategy achieves a superior balance between stealth and attack effectiveness.

The main contributions of this paper are summarized as follows:
\begin{itemize}
    \item We propose a novel cognitive distribution-driven model extraction mechanism that incorporates attention decay theory into black-box recommendation model distillation. This advances the extraction process from discrete ranking to continuous value modeling,  leading to significantly improved surrogate model accuracy and semantic fidelity.
    \item We develop a pollution item generation framework that ensures behavioral pattern consistency. By combining collaborative relation analysis with gradient-based optimization, our method produces attack sequences that are both natural and effective, ensuring high stealth without compromising attack potency.
    \item 
    Extensive experiments on multiple public datasets demonstrate that our approach significantly outperforms existing baselines in both attack performance and evasion rate, underscoring the value of cognitive modeling and behavioral synergy in building secure recommender systems.
\end{itemize}

\section{Related Work}
\subsection{Model Extraction in Recommendation}
Model Extraction Attacks aim to construct a functionally equivalent or similar surrogate model through limited API queries, without access to the target model’s internal parameters. Early efforts targeted image and text classification \cite{tramer2016stealing}, such as Knockoff nets \cite{orekondy2019knockoff} and MAZE \cite{kariyappa2021maze}. More recently, tailored extraction methods for recommender systems have emerged. DFME \cite{yue2021black} pioneered a data-free framework, distilling surrogates via pairwise ranking loss from synthetic user behavior sequences. FSME \cite{zhang2024few} extended DFME by incorporating limited genuine data for improved efficiency. Furthermore, ME-MIA \cite{zhu2023membership} adapted this framework for membership inference attacks, using a hybrid BPR+Hinge loss.

However, current black-box extraction methods relying on discrete relative order discrepancies overlook continuous value differences and inherent attention decay in user perception. This results in sparse distillation signals, significant information loss, and often leads to unstable or overfitted surrogate models, hindering effective transfer attacks. Our approach, in contrast, employs continuous value distillation to accurately recover target system preferences, providing a robust, transferable knowledge base for subsequent attacks.

\subsection{Attacks on Recommender Systems}
Profile pollution attacks \cite{lam2004shilling} pose a significant adversarial threat to recommender systems, involving attackers injecting crafted items into user behavior sequences to manipulate recommendation outcomes. Previous approaches relied on heuristic strategies, e.g., inserting random high-frequency items, which often failed to account for model internals and yielded limited attack efficacy \cite{xing2013take}. Recent advances have shifted towards gradient-guided methods \cite{yue2021black}, which identify optimal pollution items in the embedding space for improved precision. For instance, INFAttack \cite{du2024precision} introduced an influence function within a gradient-based framework to generate more efficient pollution sequences.

However, gradient-based methods often select items solely by embedding proximity, neglecting real-world semantic relevance. This leads to behaviorally inconsistent sequences easily detectable by modern defenses. To mitigate this, our novel pollution generation strategy integrates collaborative relationship analysis with gradient optimization, uniquely preserving natural user behavior patterns while ensuring strong target item promotion and operational stealth.

\section{Methodology}
\label{sec:method}
\begin{figure}[tp]
  \centering
    \includegraphics[scale=0.45]{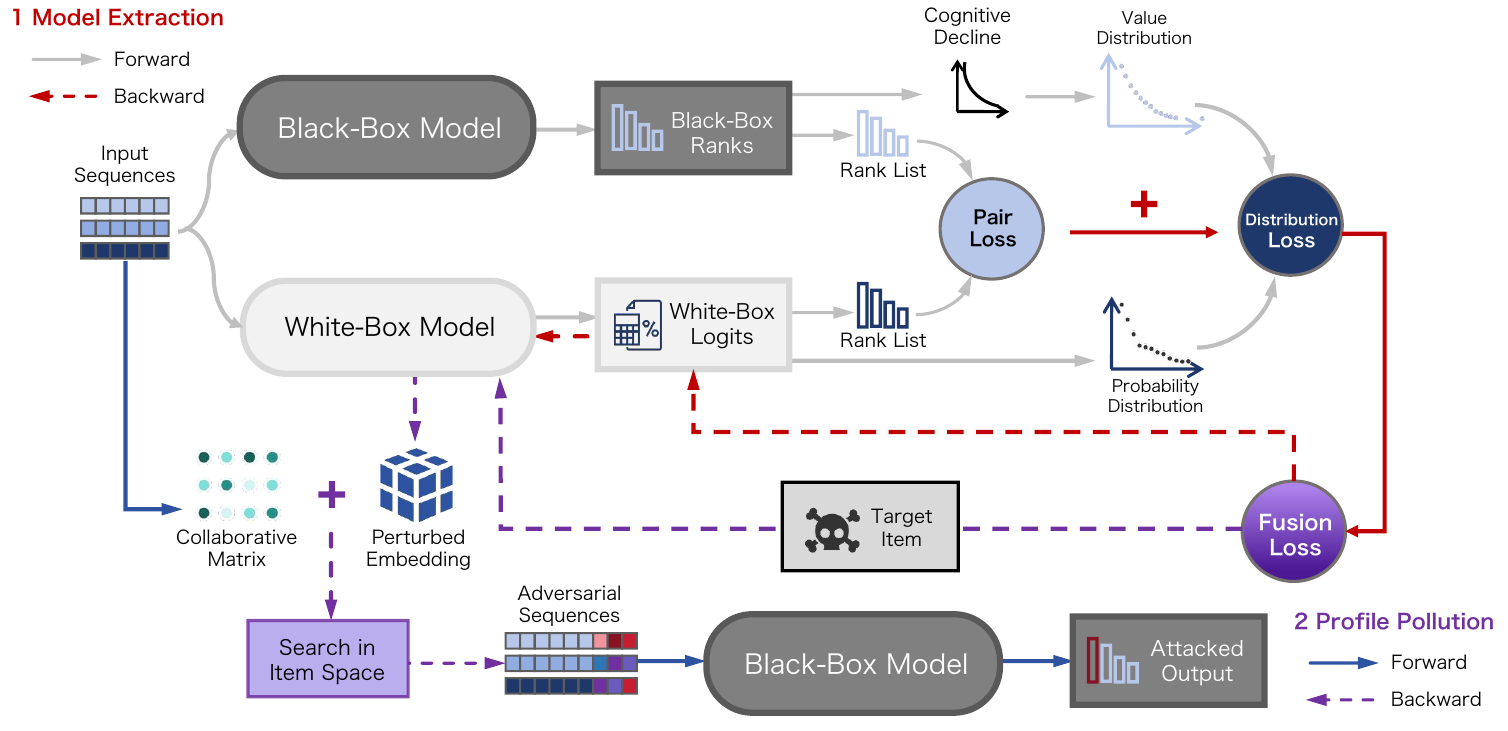}

  \caption{Dual-enhanced black-box attack framework.}
  \label{fig-framework}
\end{figure}
\label{sec:overview}
We propose a dual-enhanced black-box attack framework, shown in Fig.~\ref{fig-framework}, encompassing two core stages: (i) \textbf{cognitive distribution-aligned distillation} for learning a surrogate model that faithfully recovers ranking preferences from limited black-box queries; and (ii) \textbf{behavior-consistent pollution item generation}, achieved by a dual-signal fusion integrating gradient guidance with historical collaborative relations. This pipeline, from cognitive modeling and surrogate training to attack generation and promotion validation, ensures both stealthiness and efficacy in sequential recommendation systems.

\subsection{Problem Definition and Notations}
\label{sec:problem}
Let  \( f_b: \mathcal{X} \rightarrow \mathcal{I}^k \) be the target black-box recommender, where \(\mathcal{X}\) is the user sequence space and \( \mathcal{I}^k \) is an ordered top-\(k\) list. A user sequence $x = [i_1, i_2, \ldots, i_T]$ has length \(T\), with $ i_t \in \mathcal{I}$ over the item universe \(\mathcal{I}\). The attacker, lacking \(f_b\)'s parameters or training data, observes only truncated rankings via limited API queries. Our objective is learn a high-fidelity surrogate \( f_w \) and generate an injection sequence \( z \) effectively promote a target item \( i^* \) in \( f_b \).

During distillation, we collect a query set $\mathcal{D} = \{(x_i, I_i^k)\}_{i=1}^{B}$, where  $I_i^k = f_b(x_i)$. Surrogate $f_w$ is then optimized by minimizing distillation loss 
$\mathcal{L}_{\mathrm{dis}}$:

\begin{equation}
f_w^* = \arg\min_{f_w} \sum_{i=1}^{B} \mathcal{L}_{\mathrm{dis}} \big(f_w(x_i), I_i^k \big).
\end{equation}

For attack generation, given a collaborative matrix \( \mathbf{S} \) (from historical logs) and surrogate gradients \( \nabla_x \mathcal{L}_{\mathrm{atk}} \), we seek an optimal injection sequence $z$
to maximize target item \( i^* \)'s expected exposure under \( f_b \):
\begin{equation}
z^* = \arg\max_{z} \, E_t \big( f_b([x; z]) \big) 
\quad \text{s.t. } z \sim \text{Dual-Enhanced}(f_w, \mathbf{S}).
\end{equation}

\subsection{Data Generation and Model Distillation}
\label{sec:distillation}

\subsubsection{Data Generation}
We synthesize query sequences via an autoregressive sampling strategy. Starting with a random seed item, we repeatedly query the black-box API for a top-\(k\) list and sample the next item using a prescribed policy, e.g., position-decay or temperature sampling, until a preset length or budget is reached. This generates sequences reflecting real interaction temporal dependencies, within a limited query budget.

\subsubsection{Model Distillation}
Our surrogate aims to reproduce the black-box model's rankings and capture its implicit attention distribution pattern. Inspired by cognitive psychology's Primacy Effect and position bias, users' attention and perceived value for list items decay exponentially with ranking position. These insights drive our attention-decay mechanism, evolving distillation from order alignment to cognitive distribution alignment.

Specifically, for an input sequence $x$, the black-box model outputs an ordered top-$k$ list: $I^k(x) = [i_1, i_2, \ldots, i_k]$. Assuming user attention intensity decays exponentially with position, we define this cognitive prior: $v(j) = \alpha^{j-1}$ with $ \alpha \in (0,1)$ controlling decay rate  (smaller $\alpha$ for stronger focus on top-ranked items). With temperature parameter $\tau_b > 0$, this prior converts to a soft black-box distribution: 
\begin{equation}
p_b(i_j \mid x) = \frac{\exp(v(j) / \tau_b)}{\sum_{t=1}^{k} \exp(v(t) / \tau_b)}.
\end{equation}
This reconstructs a continuous attention distribution from a discrete top-$k$ ranking. Unlike traditional order-mimicking distillation, our approach better aligns with the user perception's gradient nature, providing a differentiable signal. Our attention-decay formulation is also theoretically consistent with NDCG-like position discounting. Specifically, while NDCG uses $d(i) = 1/\log_2(j+1)$,  our cognitive value function is $v(j) = \alpha^{j-1}$. An $\alpha \in (0,1)$ ensures $v(j)$ and $d(j)$ are monotonic and rank-equivalent: $d(j_1) > d(j_2) \iff v(j_1) > v(j_2), \forall\, j_1 < j_2$. Thus, our attention-based cognitive distribution smoothly approximates the NDCG discounting mechanism, consistent with classical metrics.

For example, a black-box ranking of A, B, C (1st, 2nd, 3rd) is treated as a discrete ordering by traditional distillation (A $>$ B $>$ C). Our method, however, assigns attention weights (e.g., 0.6, 0.3, 0.1) to learn the attention intensity distribution. The surrogate model then produces scores $s_w(i_j; x)$ per position, normalized by a temperature $\tau_w$ into $p_w(\cdot \mid x; \tau_w)$. For ensuring cognitive-level consistency beyond rank imitation, we measure the distribution discrepancy using Kullback–Leibler divergence:
\begin{equation}
\mathcal{L}_{\mathrm{KL}}(x) = \mathrm{KL}\!\left( p_b(\cdot \mid x) \,\|\, p_w(\cdot \mid x; \tau_w) \right).
\end{equation}

However, distribution alignment alone can compromise local ranking order. We thus incorporate a pairwise distillation loss, inspired by prior work, to maintain local structural stability:
\begin{equation}
\resizebox{0.9\linewidth}{!}{$
\mathcal{L}_{\mathrm{pair}}(x) 
= \frac{1}{k-1}\sum_{j=1}^{k-1} \max\!\left\{0,\, s_w(i_{j+1}) - s_w(i_j) + \delta_1\right\}
+ \frac{1}{k}\sum_{j=1}^{k} \max\!\left\{0,\, s_w(i_j^{\mathrm{neg}}) - s_w(i_j) + \delta_2\right\},
$}
\end{equation}
where $i_j^{\mathrm{neg}}$ denotes a negative sample paired with $i_j$, and $\delta_1, \delta_2 > 0$ are margin parameters. This loss ensures local ranking consistency and discriminability while enabling global cognitive alignment. 

The final objective combines both global and local signals as follows. Here, $\lambda$ balances structure preservation: larger $\lambda$ prioritizes fine-grained ranking for precise distillation; smaller $\lambda$ emphasizes global cognitive imitation and user attention.
\begin{equation}
\mathcal{L}_{\mathrm{distill}}(x) = 
\lambda \, \mathcal{L}_{\mathrm{pair}}(x) + (1 - \lambda) \, \mathcal{L}_{\mathrm{KL}}(x),
\quad \lambda \in [0,1].
\end{equation}

\begin{figure}[tp]
  \centering
  \includegraphics[width=1\textwidth, page=1]{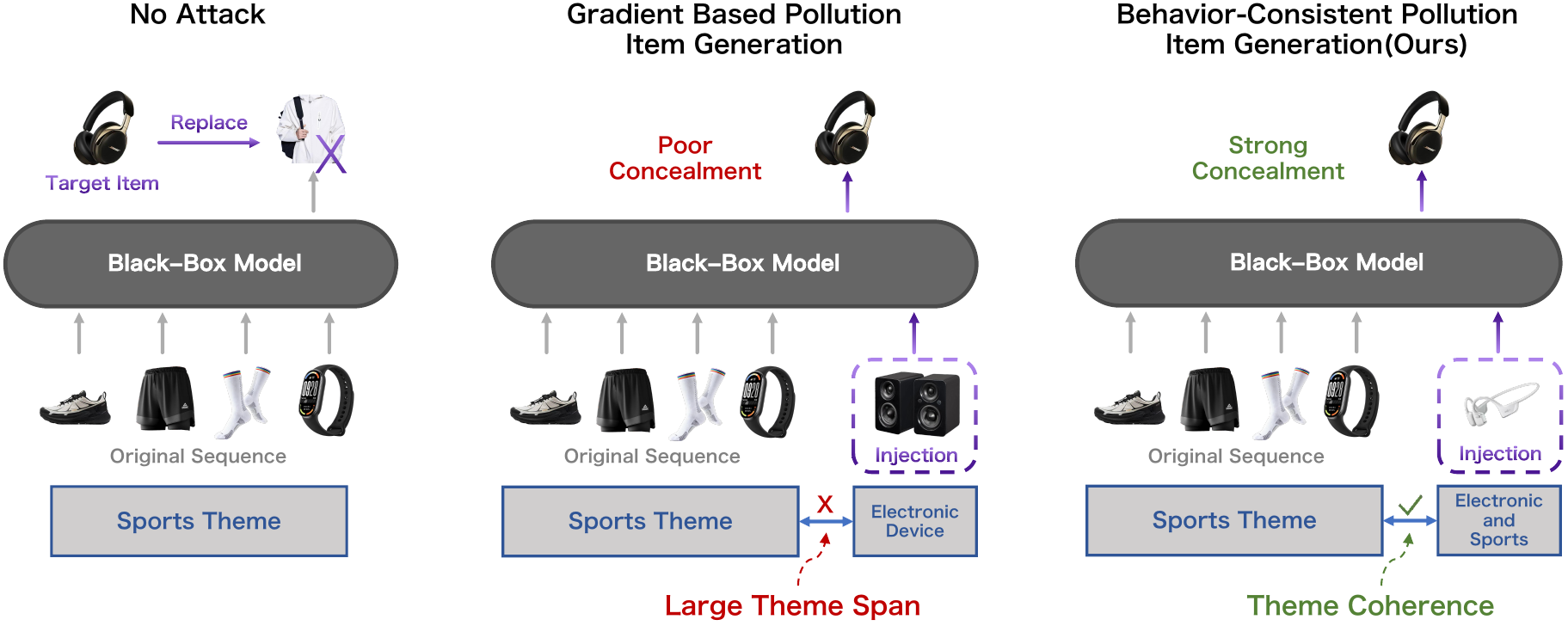}
  \caption{Concealment comparison: gradient-based vs our behavior-consistent pollution. Pure gradient attacks select semantically unrelated items (e.g., an intelligent speaker for running headphones), improving ranking but violating user patterns and increasing detectability. Our method achieves higher concealment by introducing items (e.g., object motion headphones) aligning with both gradient effectiveness and historical collaborative behavior.}
  \label{fig-profile-attack}
\end{figure}

\subsection{Behavior-Consistent Pollution Item Generation}
\label{sec:pollution}
We generate injection sequences using a behavior-consistent pollution item generation (Fig.~\ref{fig-profile-attack}) that integrates \emph{gradient guidance} and \emph{behavioral collaboration} to ensure both promotion strength and behavioral plausibility.


\begin{enumerate}
    \item \textbf{Candidate Pool via Collaborative Prior.} From historical logs, we construct a collaborative matrix \( \mathbf{S} \) and compute a conditional collaboration score \( \tilde{s}(j \mid i^*) \) for candidate item \( j \) w.r.t. target item \( i^* \). Items with high \( \tilde{s}(j \mid i^*) \) constitute a semantically consistent candidate pool.
    \item \textbf{Gradient-Guided Scoring.} We define an attack loss \( \mathcal{L}_{\mathrm{atk}} \) on the surrogate and compute the input-embedding gradient \( g = \nabla_{x} \mathcal{L}_{\mathrm{atk}} \). For candidate \( j \) with embedding \( \mathbf{e}_j \), we measure its gradient alignment \( \mathrm{sim}_g(j) \), typically via cosine similarity with the negative averaged gradient direction.
    \item \textbf{Dual-Signal Fusion.} The final candidate score $S(i)$ for item $j$ linearly combines gradient alignment $\mathrm{sim}_g(j)$ and collaborative score $\tilde{s}(j \mid i^*)$:
 
    \begin{equation}
    \mathrm{S}(j) = w_g \cdot \mathrm{sim}_g(j) + w_s \cdot \tilde{s}(j \mid i^*),
    \quad w_g, w_s \ge 0, \; w_g + w_s = 1.
    \end{equation}
    \item \textbf{Sequence Construction and Validation.} Top-ranked candidates are selected, and a greedy or beam search assembles the injection sequence within budget. This sequence is then validated on \( f_b \) and limited refinements are made if necessary.
\end{enumerate}

This mechanism jointly optimizes directional effectiveness and behavioral consistency, balancing stealthiness and promotion performance. Details are in Algorithm~\ref{alg:dual-signal}.

\begin{algorithm}[t]
\caption{Behavior-Consistent Profile Pollution}
\label{alg:dual-signal}
\begin{algorithmic}[1]
\Require sequence $\mathbf{x}$, target item $t$, expected polluted length $T$, white-box model $\mathbf{f}_{w}$ (i.e., $\mathbf{f}_{w,e}$, $\mathbf{f}_{w,m}$), step size $\epsilon$, candidates per step $n$, mixing weight $\lambda \in [0,1]$, neighbor size $K$
\Ensure polluted sequence $\mathbf{z}$
\State $\mathbf{z} \gets \mathbf{x}$

\State Precompute collaborative neighbors for $t$:
\Statex \hspace{\algorithmicindent}$\mathcal{N}_t \gets \text{Top-}K \ \arg\max_{i \in \mathcal{I}} \ \mathrm{CoRel}(i, t)$
\Comment{e.g., co-view/co-purchase PMI, Jaccard, or category proximity}

\While{\text{length of } $\mathbf{z} < T$}
  \State Append target as a placeholder: $\mathbf{z} \gets [\mathbf{z}; t]$
  \State Encode sequence: $\tilde{\mathbf{z}} \gets \mathbf{f}_{w,e}(\mathbf{z})$
  \State Backprop w.r.t.\ the placeholder using CE loss: 
  $\nabla \gets \nabla_{\tilde{\mathbf{z}}}\, \mathcal{L}_{\mathrm{ce}}(\mathbf{f}_{w,m}(\tilde{\mathbf{z}}), t)$
  \State Gradient-guided embedding: $\hat{\mathbf{z}} \gets \tilde{\mathbf{z}} - \epsilon \, \sign(\nabla)$
  \State Gradient signal: ${\mathrm{sim}_g(j)} \gets \cosSim\!\big(\hat{\mathbf{z}}, \mathbf{f}_{w,e}(i)\big)$ for all $i \in \mathcal{I}$
  \State Collaborative signal: ${\tilde{s}(j \mid i^*)} \gets \mathrm{CoRel}(i,t)$ for all $i \in \mathcal{I}$
  \State Cohort filter (keep plausible items): 
  $\mathcal{H} \gets \mathcal{N}_t \ \cup \ \text{Top-}K\ \arg\max_i \ {\mathrm{sim}_g(j)}$
  \State Dual-Signal fusion: $S(i) \gets w_g \cdot \mathrm{sim}_g(j) + w_s \cdot \tilde{s}(j \mid i^*)$ for $i \in \mathcal{H}$
  \State Pick $n$ best candidates: $\mathbf{C} \gets \text{Top-}n\ \arg\max_{i \in \mathcal{H}} S(i)$ \ \ (exclude $t$ if repeated)
  \State Instantiate candidate and keep the best:
  \Statex \hspace{\algorithmicindent} $\mathbf{z}^{(c)} \gets \mathbf{z}$ with the last $t$ replaced by $c$, for each $c \in \mathbf{C}$
  \Statex \hspace{\algorithmicindent} choose $\mathbf{z} \gets \arg\max_{\mathbf{z}^{(c)}} \ \mathbf{f}_{w}(\mathbf{z}^{(c)})_{[t]}$
  \Comment{optional: break ties by a plausibility score}
\EndWhile

\State \textbf{return} $\mathbf{z}$
\end{algorithmic}
\end{algorithm}

\section{Experiments}
In this section, we conduct extensive experiments to address the following research questions:
\begin{itemize}
    \item \textbf{RQ1:} What does the proposed cognitive distribution-driven knowledge extraction mechanism outperform existing baseline methods in the general model extraction task?
    \item \textbf{RQ2:} How effective is the proposed behavioral pattern-consistent data poisoning strategy in successfully promoting the rank of the target item?
\item  \textbf{RQ3:}  What are the individual contributions of the cognitive distribution alignment mechanism and the behavioral pattern-consistent generation strategy to the overall performance of the attack framework, as revealed by ablation studies?
\item \textbf{RQ4:} How sensitive are the model extraction and attack outcomes to the key hyper-parameters?
\item  \textbf{RQ5:} Do the generated poisoning sequences exhibit statistical indistinguishability from genuine user behavior sequences in terms of behavioral flow consistency and semantic coherence, thereby ensuring the stealthiness of the attack?
\end{itemize}

\subsection{ Experimental Setup}

\subsubsection{Dataset}
For the experimental evaluation of our proposed framework, we utilize three publicly available datasets: Movielens-1M (ML-1M) \cite{harper2015movielens}, Steam \cite{mcauley2015image}, and Amazon Beauty \cite{ni2019justifying}. As shown in Table \ref{tab.1}, these datasets exhibit diverse characteristics, which allow for a comprehensive assessment of our framework's generality and robustness.

\subsubsection{Baselines}
To comprehensively evaluate our framework, we compare it against representative baselines in two aspects:  

\begin{enumerate}
    \item \textbf{Model Extraction Attacks}
    \begin{itemize}
        \item \textbf{DFME}\cite{yue2021black}: Extracts black-box models via synthetic sequence generation and pairwise ranking distillation, aligning the order of top‑k items through marginal ranking loss.  
        \item \textbf{ME-MIA} \cite{zhu2023membership}: Optimizes extraction using a combined BPR ranking loss and Hinge loss, achieving enhanced performance via engineering refinements.  
    \end{itemize}
    \item \textbf{Profile Pollution Attacks}
    \begin{itemize}
        \item \textbf{RandAlter}\cite{song2020poisonrec}: Injects random items alternating with target items. 
        \item  \textbf{DQN} \cite{zhang2020practical,zhao2017deep}: Employs deep Q-learning with RNN to maximize target item ranking.  
        \item \textbf{WhiteBox SimAlter} \cite{yue2021black}: Inserts items similar to the target based on white-box embeddings. 
        \item \textbf{DFME-based pollution} \cite{yue2021black}: Leverages a distilled white-box model and gradient signals to craft adversarial sequences that promote target exposure.  
    \end{itemize}
\end{enumerate}

All experiments are conducted using three widely used sequential recommendation models as surrogate architectures: (1) \textbf{NARM} \cite{li2017neural}  (RNN-based with attention), \textbf{SASRec} \cite{kang2018self} (self-attentive Transformer), and \textbf{BERT4Rec} \cite{sun2019bert4rec} (bidirectional Transformer with masked modeling). These cover a spectrum from local dependency modeling to long-range contextual learning.

\begin{table}[t]
\centering
\caption{Detailed statistics of experimental datasets.}\label{tab.1}
\begin{tabular}{c|c|c|c|c|c}
\hline
Datasets & Users & Items & Averag length & Max length & Sparisity \\
\hline
ML-1M &  6,040
 & 3,416 & 166 & 2,277 & 95.16\%\\
Steam &  334,542
 & 13,046 & 13 & 2,045 & 99.90\%\\
Beauty &  40,226
 & 54,542 & 9 & 293 & 99.98\%\\
\hline
\end{tabular} 
\end{table}
\subsubsection{Evaluation Metrics}
To evaluate our framework, we employ standard ranking metrics (Recall@K and NDCG@K) to assess recommendation quality, along with Agreement@K (Agr@K) to measure consistency between the black-box target and extracted white-box models, following established protocols \cite{yue2021black,sun2019bert4rec,kang2018self}.


\subsubsection{Implementation Details}

Our experiments follow standard sequential recommendation protocols. For each user sequence $\bm{x}$ of length $T$, we use $\bm{x}_{[:T-2]}$ for training, the penultimate item for validation, and the last item for testing. To ensure fair comparison, we adhere to the hyperparameter settings from baseline methods [10,14,23], employing Adam optimizer (lr=0.001, weight decay=0.01, batch size=128) with 100  warmup steps. Maximum sequence lengths are set to 200 (ML-1M) and 50 (Steam, Beauty).

Model architectures include: NARM (1 GRU layer), SASRec, and BERT4Rec (2 Transformer layers, 2 attention heads). Our framework uses $\alpha=0.97$ for cognitive distribution attenuation and temperature=0.5. Dataset-specific parameters (dropout: 0.1-0.5; BERT4Rec masking: 0.2-0.6) follow original publications. For attack evaluation, we set profile pollution sizes to 1\% of sequence lengths (10 items for ML-1M, 2 for Steam/Beauty) with poisoned user counts of 60 (ML-1M), 3345 (Steam), and 402 (Beauty) \cite{yue2021black}


\subsection{Model Extraction(RQ1)}

To rigorously evaluate our framework, we conducted comprehensive experiments across three benchmark datasets (ML-1M, Steam, and Beauty) from three critical perspectives. 
 The specific results, as shown in Table \ref{tab1}, Table \ref{tab2}, and Table \ref{tab3}, systematically demonstrate the effectiveness of our method. 

\begin{itemize}
    \item \textbf{Within-Model Comparison} 

    Using NARM on ML-1M as an example, our method achieves superior performance on both recommendation quality and ranking consistency, significantly outperforming DFME and MIA baselines. This validates that our cognitive distribution-driven distillation better captures the original model's ranking preferences, especially in critical top positions.
\item  \textbf{ Cross-Architecture Transferability}

Our method consistently achieves the best Agr@1 and Agr@10 metrics across all nine model-dataset combinations. The advantage is particularly pronounced for complex architectures like BERT4Rec, where Agr@1 improves by over 20\% compared to baselines. Notably, while traditional methods degrade when moving from RNN-based to Transformer-based architectures, our approach maintains stable performance, demonstrating strong adaptability to different model internal structures.

\begin{table}[t]
\centering
\caption{Performance comparison of different  model extraction  methods on ML-1M dataset.}\label{tab1}
\renewcommand{\arraystretch}{1.2}
\setlength{\tabcolsep}{8pt}
\begin{tabular}{c|c|c|c|c|c}
\hline
Method & Model Extraction & N@10 & R@10 & Agr@1 & Agr@10 \\
\hline
\multirow{4}{*}{NARM} 
& BlackBox & 0.626 & 0.819 & - & - \\
\cline{2-6}
& WhiteBox-DFME & 0.615 & 0.812 & 0.571 & 0.747 \\
& WhiteBox-ME-MIA & 0.615 & 0.811 & 0.551 & 0.724 \\
& \textbf{WhiteBox-Ours} & \textbf{0.617} & \textbf{0.813} & \textbf{0.601} & \textbf{0.757} \\
\hline
\multirow{4}{*}{SASRec} 
& BlackBox & 0.605 & 0.827 & - & - \\
\cline{2-6}

& WhiteBox-DFME & \textbf{0.602} & 0.802 & 0.454 & 0.662 \\
& WhiteBox-ME-MIA & 0.594 & 0.820 & 0.528 & 0.711 \\
& \textbf{WhiteBox-Ours} & 0.595 & \textbf{0.818} & \textbf{0.573} & \textbf{0.724} \\
\hline
\multirow{4}{*}{BERT4Rec} 
& BlackBox & 0.590 & 0.794 & - & - \\
\cline{2-6}

& WhiteBox-DFME & 0.559 & 0.775 & 0.335 & 0.581 \\
& WhiteBox-ME-MIA & 0.560 & 0.773 & 0.308 & 0.569\\
& \textbf{WhiteBox-Ours} & \textbf{0.571} & \textbf{0.782} & \textbf{0.409} & \textbf{0.631} \\
\hline
\end{tabular}
\end{table}

\begin{table}[t]
\centering
\caption{Performance comparison of different model extraction methods on the Steam dataset.}\label{tab2}
\renewcommand{\arraystretch}{1.2}
\setlength{\tabcolsep}{8pt}
\begin{tabular}{c|c|c|c|c|c}
\hline
Method & Model Extraction & N@10 & R@10 & Agr@1 & Agr@10 \\
\hline
\multirow{4}{*}{NARM} 
& BlackBox & 0.630 & 0.848 & - & - \\
\cline{2-6}

& WhiteBox-DFME & 0.601 & 0.806 & 0.743 & 0.722 \\
& WhiteBox-ME-MIA & 0.612 & 0.821 & 0.728 & 0.752 \\
& \textbf{WhiteBox-Ours} & \textbf{0.621} & \textbf{0.833} & \textbf{0.760} & \textbf{0.763} \\
\hline
\multirow{4}{*}{SASRec} 
& BlackBox & 0.617 & 0.839 & - & - \\
\cline{2-6}

& WhiteBox-DFME & 0.593 & 0.805 & 0.668 & 0.702 \\
& WhiteBox-ME-MIA & 0.602 & 0.826 & 0.639 & 0.767 \\
& \textbf{WhiteBox-Ours} & \textbf{0.615} & \textbf{0.834} & \textbf{0.740} & \textbf{0.782} \\
\hline
\multirow{4}{*}{BERT4Rec} 
& BlackBox & 0.627 & 0.851 & - & - \\
\cline{2-6}

& WhiteBox-DFME & 0.585 & 0.793 & 0.708 & 0.607 \\
& WhiteBox-ME-MIA & 0.598 & 0.811 & 0.723 & 0.695 \\
& \textbf{WhiteBox-Ours} & \textbf{0.613} & \textbf{0.827} & \textbf{0.744} & \textbf{0.721} \\
\hline
\end{tabular}
\end{table}

\begin{table}[t]
\centering
\caption{Performance comparison of different model extraction methods on Beauty dataset.}\label{tab3}
\renewcommand{\arraystretch}{1.2}
\setlength{\tabcolsep}{8pt}
\begin{tabular}{c|c|c|c|c|c}
\hline
Method & Model Extraction & N@10 & R@10 & Agr@1 & Agr@10 \\
\hline
\multirow{4}{*}{NARM} 
& BlackBox & 0.356 & 0.517 & - & - \\
\cline{2-6}

& WhiteBox-DFME & 0.261 & 0.370 & 0.294 & 0.381 \\
& WhiteBox-ME-MIA & 0.271 & 0.378 & 0.267 & 0.369 \\
& \textbf{WhiteBox-Ours} & \textbf{0.282} & \textbf{0.385} & \textbf{0.349} & \textbf{0.468} \\
\hline
\multirow{4}{*}{SASRec} 
& BlackBox & 0.341 & 0.493 & - & - \\
\cline{2-6}

& WhiteBox-DFME & 0.333 & 0.483 & 0.433 & 0.423 \\
& WhiteBox-ME-MIA & 0.334 & 0.486 & 0.413 & 0.424 \\
& \textbf{WhiteBox-Ours} & \textbf{0.337} & \textbf{0.489} & \textbf{0.439} & \textbf{0.433} \\
\hline
\multirow{4}{*}{BERT4Rec} 
& BlackBox & 0.321 & 0.493 & - & - \\
\cline{2-6}

& WhiteBox-DFME & 0.191 & 0.269 & 0.351 & 0.342 \\
& WhiteBox-ME-MIA & 0.190 & 0.267 & 0.353 & 0.345 \\
& \textbf{WhiteBox-Ours} & \textbf{0.191} & \textbf{0.278} & \textbf{0.365} & \textbf{0.362} \\
\hline
\end{tabular}
\end{table}

\item \textbf{Cross-Dataset Robustness}

     In data-rich scenarios (ML-1M and Steam), our method preserves recommendation quality while achieving the highest ranking consistency. In sparse environments (Beauty), the relative advantage becomes more pronounced, with Agr@10 improving by 23\% over DFME.  These observations yield an important insight: in sparse data environments, conventional techniques relying on hard labels or pairwise learning exhibit heightened sensitivity to data noise, whereas our cognitive distribution modeling effectively dampens such interference through its attention decay mechanism, thereby demonstrating superior robustness and generalization capability.

\end{itemize}


\subsection{Profile  Pollution attack  Attack(RQ2)}

\begin{figure}[t] 
  \centering
  \includegraphics[width=1\textwidth, page=1]{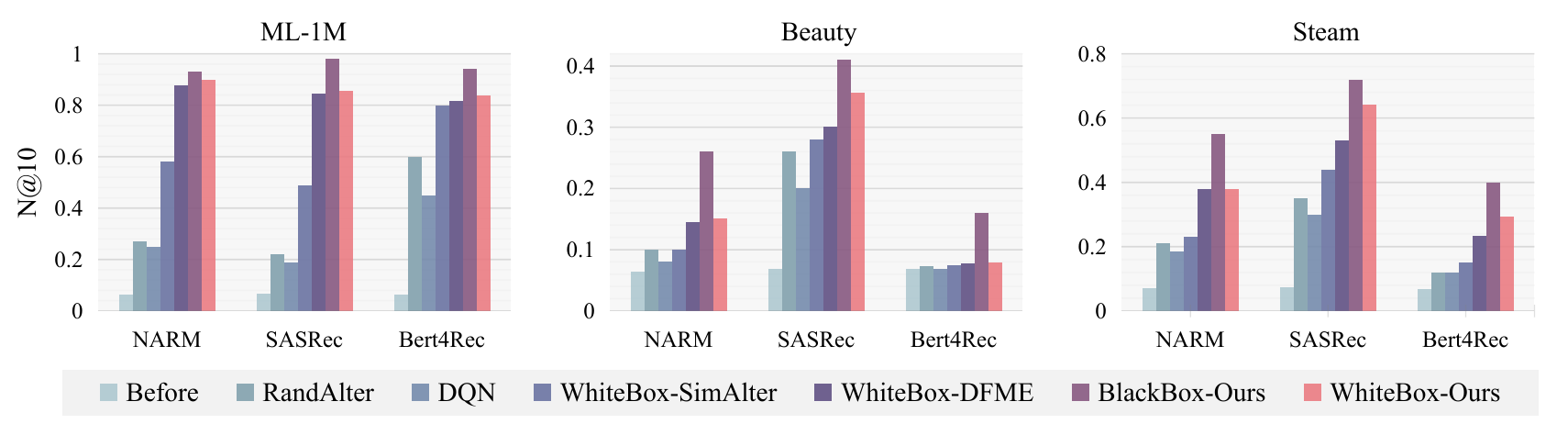}
  \caption{Performance comparison of different profile pollution attack methods.}
  \label{figattackmain}
\end{figure}

The proposed behavioral pattern-consistent poisoning strategy demonstrates significant effectiveness in promoting target item rankings across diverse experimental settings. The results are shown in Fig. \ref{figattackmain}.  On the ML-1M dataset, our method achieves optimal performance under both white-box and black-box configurations, with the WhiteBox-Ours version improving target item hit rates by approximately 2.4\% over the strongest baseline. This advantage proves particularly pronounced in complex architectures like BERT4Rec, while maintaining consistent superiority across different model types. The strategy also exhibits remarkable robustness in sparse data environments like the Beauty dataset, where it sustains leading performance despite overall metric degradation across all methods.

Comprehensive evaluation across three datasets confirms the method's stable and efficient attack capability under varying model architectures and data densities. The core innovation of synergistic collaboration analysis and gradient optimization enables the generated sequences to preserve semantic coherence with genuine user behavior while effectively elevating target item rankings. Particularly in black-box scenarios, attack success rates reach from 0.93 to 0.98, demonstrating an optimal balance between attack potency and behavioral stealth through precise gradient guidance.

\subsection{Ablation Studies(RQ3)}
\begin{figure}[t]
  \centering
    \includegraphics[scale=0.4]{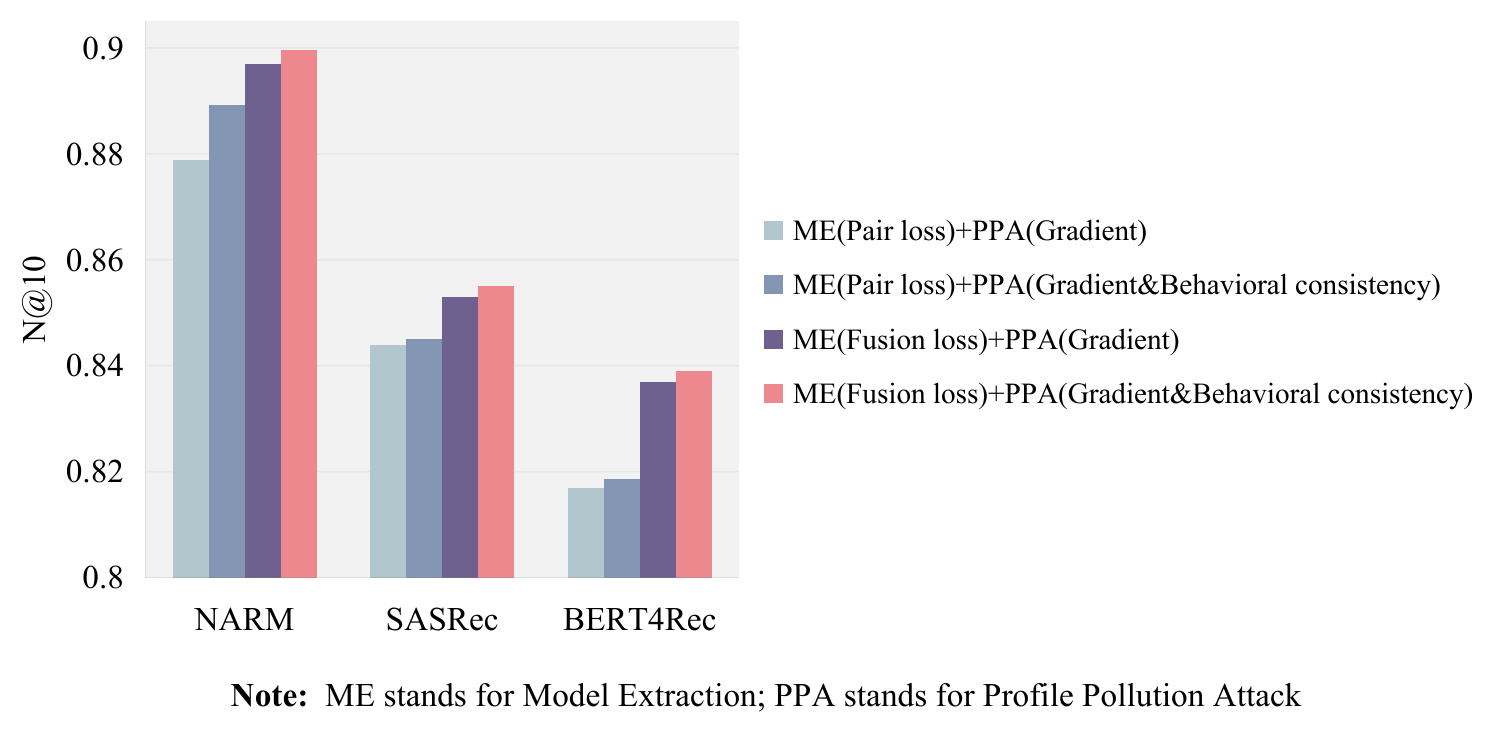}
  \caption{Results of ablation experiments on the ML-1M dataset.}
  \label{figmelt}
\end{figure}

The ablation experiment results as shown in Fig. \ref{figmelt} demonstrate that the cognitive distribution alignment mechanism is the core factor in improving attack effectiveness. In the BERT4Rec model, using the fusion loss alone significantly improved the attack success rate from 0.817 to 0.837, an improvement that far exceeds that achieved by adding the behavioral consistency strategy alone (0.817 to 0.8187). This proves that cognitive distribution alignment can more effectively capture ranking semantic information.
Although the behavioral consistency strategy shows limited improvement when applied alone, it exhibits significant effects when combined with cognitive distribution alignment. In the NARM model, their combination achieved optimal performance (0.8996). This synergistic effect indicates that cognitive distribution alignment ensures accurate acquisition of model ranking knowledge, while behavioral consistency guarantees attack stealth, with their organic combination forming the key foundation for our method's superiority.

\subsection{Impact of Hyper-parameters  (RQ4)}
\begin{figure}[t]

  \centering
  \includegraphics[width=1\textwidth, page=1]{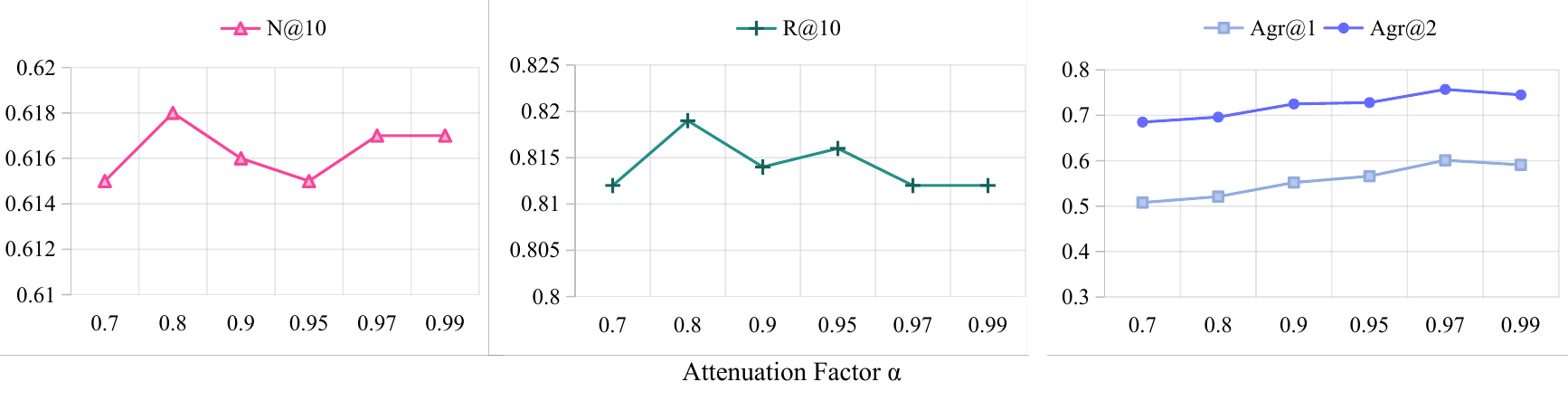}
  \caption{Comparison of Different Attenuation Factors Extracted by NARM Structure on ML1M Dataset.}
  \label{fig-alpha}
\end{figure}

We conducted systematic parameter sensitivity experiments to validate the effect of the attenuation factor $\alpha$ in the cognitive distribution mechanism. As shown in Fig. \ref{fig-alpha}, as $\alpha$ increases from 0.7 to 0.97, the ranking consistency metrics Agr@1 and Agr@10 show a steady upward trend, peaking at 0.601 and 0.757, respectively, at $\alpha$=0.97, representing approximately 18\% and 10\% improvements compared to $\alpha$=0.7. This pattern confirms our theoretical hypothesis: higher $\alpha$ values (closer to 1) better align with the cognitive patterns of primacy effect and position bias, enabling more accurate capture of the essential characteristics of ranking distributions. Notably, when $\alpha$ exceeds 0.97, all metrics show slight degradation, indicating an optimal parameter range (0.95 - 0.97). This finding provides important guidance for parameter setting in practical applications while simultaneously verifying the robustness of the cognitive distribution alignment mechanism to parameter selection.
 

\subsection{Case Analysis(RQ5)}

\begin{figure}[t]

  \centering
  \includegraphics[width=1\textwidth, page=1]{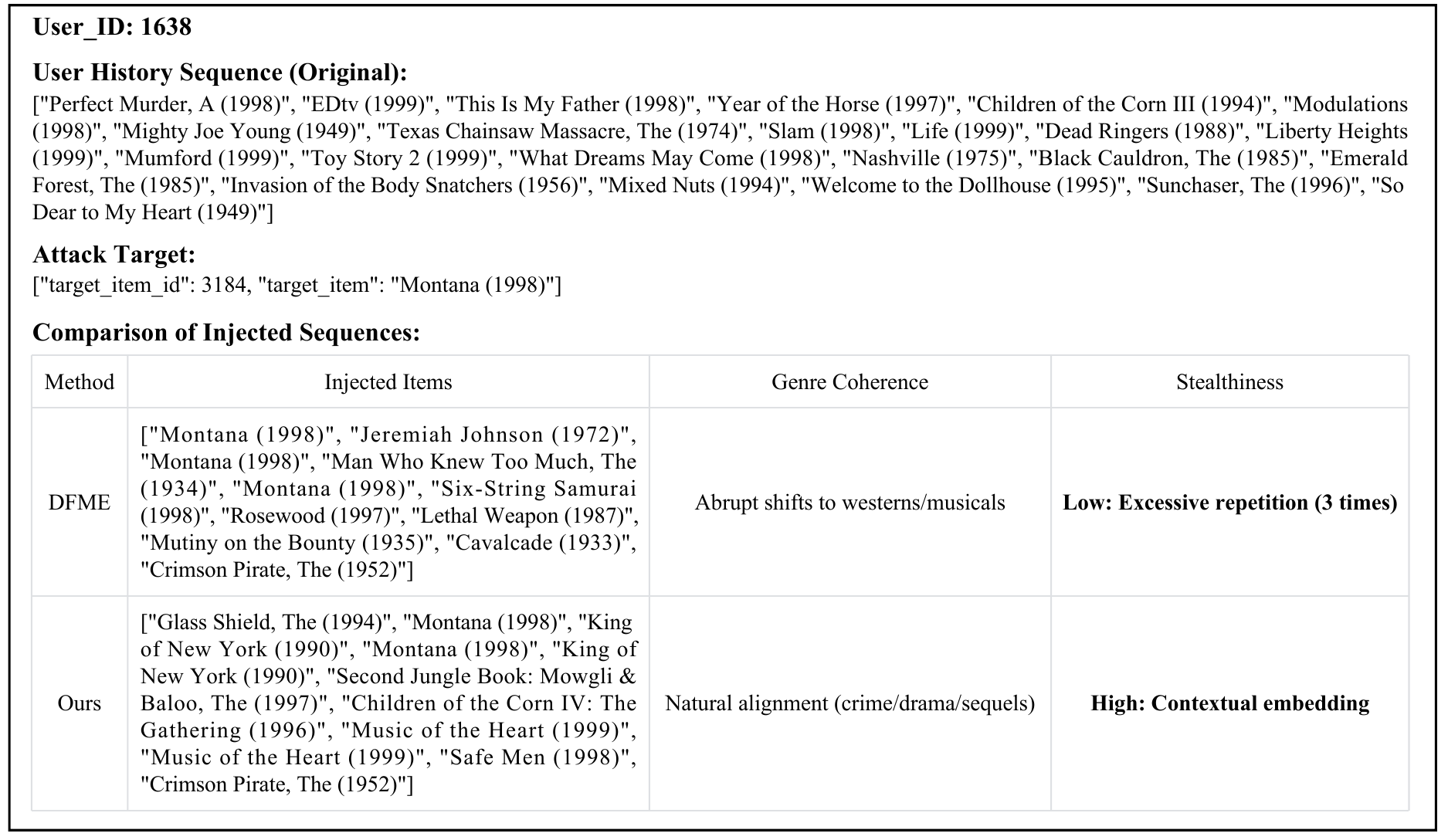}
  \caption{Comparative analysis of injected sequence cases.}
  \label{fig-case}
\end{figure}

To quantitatively validate the stealthiness of our poisoning strategy, we conduct a fine-grained case analysis on a representative user from the MovieLens dataset. This user demonstrates a clear behavioral pattern with a predominant preference for 1990s films across multiple genres, particularly showing interest in sequel movies. We compare the behavioral consistency between our method and the baseline through genre coherence and stealthiness.
As shown in Fig. \ref{fig-case}, we can find that our method maintains superior behavioral consistency by preserving the user's temporal preferences and genre interests, while naturally integrating the target item. The baseline's temporal anomalies and excessive target repetition create detectable patterns that compromise stealthiness. This demonstrates that our behavior-aware generation strategy successfully balances attack effectiveness with behavioral authenticity.

\section{Conclusion}
This work establishes that the security of sequential recommender systems requires evaluation against adversaries mimicking both the statistical and semantic patterns of real user behavior. By formalizing the alignment of cognitive value distributions and integrating behavioral consistency directly into attack optimization, our framework introduces a paradigm shift from conspicuous manipulation to imperceptible influence. Our results confirm that such cognitively grounded and behaviorally stealthy attacks pose a significantly more potent and practical threat. This necessitates a fundamental rethinking of defensive strategies, moving beyond anomaly detection based on superficial statistical deviations towards auditing the semantic plausibility and cognitive consistency of user sequences. Ultimately, our study not only exposes a critical vulnerability vector but also provides a novel analytical lens, one that integrates cognitive psychology and behavioral analysis for building next-generation, robust recommender systems secure against advanced, semantically-aware adversaries.


%
%
%

\section*{Acknowledgements}
This research was funded by the National Key Research and Development Program of China (Grant No. 2024YFC3307400), under the "Social Governance and Smart Society Technology Support" key special project.

\bibliographystyle{splncs04}
\bibliography{references}
\end{document}